\documentclass[a4paper]{article}

\usepackage{INTERSPEECH2021}
\usepackage{multirow}
\usepackage{makecell}
\usepackage{pifont}
\usepackage{amssymb}
\usepackage{graphicx}
\usepackage{subfigure}
\usepackage{cite}
\usepackage[marginal]{footmisc}

\title{Adversarial Data Augmentation for Disordered Speech Recognition}
\name{Zengrui Jin$^{1*}$, Mengzhe Geng$^{1*}$, Xurong Xie$^{2}$, Jianwei Yu$^1$, Shansong Liu$^1$, \\Xunying Liu$^1$, Helen Meng$^1$}

\address{
  $^1$The Chinese University of Hong Kong, Hong Kong SAR, China\\
  $^2$Shenzhen Institutes of Advanced Technology, Chinese Academy of Sciences, Shenzhen, China
  }
\email{\{zrjin, mzgeng, jwyu, ssliu, xyliu, hmmeng\}@se.cuhk.edu.hk, xr.xie@siat.ac.cn}

\begin{document}

\maketitle
\begin{abstract}

Automatic recognition of disordered speech remains a highly challenging task to date. 
The underlying neuro-motor conditions, often compounded with co-occurring physical disabilities, lead to the difficulty in collecting large quantities of impaired speech required for ASR system development. 
To this end, data augmentation techniques play a vital role in current disordered speech recognition systems. 
In contrast to existing data augmentation techniques only modifying the speaking rate or overall shape of spectral contour, fine-grained spectro-temporal differences between disordered and normal speech are modelled using deep convolutional generative adversarial networks (DCGAN) during data augmentation to modify normal speech spectra into those closer to disordered speech. 
Experiments conducted on the UASpeech corpus suggest the proposed adversarial data augmentation approach consistently outperformed the baseline augmentation methods using tempo or speed perturbation on a state-of-the-art hybrid DNN system. 
An overall word error rate (WER) reduction up to 3.05\% (9.7\% relative) was obtained over the baseline system using no data augmentation. 
The final learning hidden unit contribution (LHUC) speaker adapted system using the best adversarial augmentation approach gives an overall WER of 25.89\% on the UASpeech test set of 16 dysarthric speakers. 

\end{abstract}

\section{Introduction}

\let\thefootnote\relax\footnotetext{* Equal contribution was made between the first two authors.}

Despite the rapid progress of automatic speech recognition (ASR) technologies targeting normal speech task domains \cite{bahl1986maximum, deeprecurrent, Povey+2016, chan2016listen, wang2020transformer} in the past few decades, accurate recognition of disordered speech remains a highly challenging task to date \cite{christensen2013combining, sehgal2015model, yu2018development, xiong2019phonetic, xiong2020source, liu2020exploiting, wang2021improved}.
Speech disorders such as dysarthria are caused by a range of neuro-motor conditions including cerebral palsy, amyotrophic lateral sclerosis, Parkinson disease, stroke or traumatic brain injuries \cite{lanier2010speech}, leading to weakness or paralysis of muscles used in articulation and reduced intelligibility of speech for human listeners.

Disordered speech presents a range of challenges to current deep neural networks (DNNs) based ASR technologies that are predominantly targeting normal speech. 
First, a large mismatch between disordered and normal speech is often observed. 
Such difference systematically manifests themselves in articulatory imprecision, decreased volume and clarity, breathy and hoarse voice, changes in pitch, increased dysfluencies and lower speaking rate \cite{hixon1964restricted, kent2000dysarthrias}.
State-of-the-art ASR systems designed for normal speech often produce very high recognition error rate above 50\% when being applied to impaired speech \cite{wang2021improved, hermann2020dysarthric}. 
Second, the underlying neuro-motor conditions, often compounded with co-occurring physical disabilities and fatigue when speaking, lead to the difficulty in collecting large quantities of disordered speech required for ASR system development. 
For data intensive deep learning technologies widely used in current speech recognition systems, large quantities of well-matched, in-domain speech data are essential.

To this end, data augmentation techniques play a vital role to address the above data sparsity issue. 
In the context of normal speech recognition tasks, data augmentation techniques have been widely studied. 
By expanding the limited training data using, for example, tempo, vocal tract length or speed perturbation \cite{verhelst1993overlap, kanda2013elastic, jaitly2013vocal, ko2015audio}, stochastic feature mapping \cite{cui2015data}, cross domain feature adaptation \cite{bell2012transcription}, simulation of noisy and reverberated speech to improve environmental robustness \cite{ko2017study}, and more advanced generative adversarial network (GAN) based augmentation by synthesis \cite{DBLP:conf/icassp/Hu0Q18, jiao2018simulating, chen2020improving}, and end-to-end back translation in end-to-end systems \cite{hayashi2018back}, the coverage of the augmented training data and the resulting speech recognition systems’ generalization can be improved.

In contrast, so far only limited research on data augmentation approaches targeting disordered speech recognition has been conducted. 
Motivated by the spectro-temporal level differences between disordered speech and normal speech such as slower speaking rates, recent research in this direction has been largely focused on tempo-stretching \cite{vachhani2018data, xiong2019phonetic}, vocal tract length perturbation (VTLP) \cite{jaitly2013vocal}, and speed perturbation \cite{ko2015audio} of normal speech recorded from healthy control speakers. 
The resulting ``disordered like" speech carrying a slower speaking rate and modified overall vocal tract spectral shape is then used to augment the limited dysarthric speech training data. 
Alternative approaches based on cross-domain DNN adaptation \cite{christensen2013combining, yu2018development} and voice conversion \cite{wang2020end} have also been investigated. 
An investigation over a range of speech augmentation approaches for disordered speech recognition reported in \cite{geng2020investigation} suggested the combined use of speaker dependent and independent speed perturbation factors produced the largest performance improvements. 

One issue associated with the existing data augmentation approaches, for example, temporal or speed perturbation, is the lack of ability to model fine-grained differences between normal and impaired speech during the augmentation process. 
Although the overall decrease in speaking rate and speech volume as well as changes in spectral envelope can be characterized using speed perturbation, other prominent features associated with disordered speech including articulatory imprecision, decreased vocal clarity, breathy and hoarse voice as well as increased dysfluencies cannot be fully captured. 

In order to address this issue, and inspired by the recent successful application of GAN \cite{goodfellow2014generative, radford2015unsupervised} across a wide range of speech processing tasks including, but limited to, 
speech enhancement \cite{pascual2017segan, phan2020improving}, 
speaker verification \cite{DBLP:conf/interspeech/MichelsantiT17},
speech synthesis and voice conversion \cite{saito2017statistical, kameoka2018stargan}, 
code-switching sentence generation \cite{DBLP:conf/interspeech/ChangCL19}
and speech emotion recognition \cite{DBLP:conf/icassp/ChangS17}, 
speaker dependent GANs are trained to transform tempo or speed perturbed healthy speech spectra into those of individual target dysarthric speakers. 
The resulting data augmentation process not only allows the overall change in speaking rate, speech volume and spectral envelope to be simulated as in tempo or speed perturbation, but also ensures further fine-grained spectro-temporal characteristics associated with disordered speech including articulatory imprecision, decreased clarity, breathy and hoarse voice as well as increased dysfluencies and pauses to be emulated in the final augmented data. 

Experiments conducted on the largest publicly available UASpeech dysarthric speech corpus \cite{kim2008dysarthric} suggest the proposed adversarial data augmentation approach consistently outperformed the baseline augmentation methods using tempo or speed perturbation on state-of-the-art hybrid and end-to-end systems. 
An overall WER reduction up to 3.05\% (9.7\% relative) was obtained over the baseline system using no data augmentation. 
The final LHUC \cite{swietojanski2016learning}$^1$\footnote{$^1$ In order to facilitate adaptation to unseen speakers in practical system deployment, rather than building speaker dependent systems, speaker adaptively trained systems were constructed.} speaker adapted system using the best adversarial augmentation approach gave an overall WER of 25.89\% on the UASpeech test set of 16 dysarthric speakers. 
To the best of our knowledge, this is the best performance reported so far on the UASpeech task. 

The major contributions of this paper are listed below. 
To the best of our knowledge, this is the first work to systematically investigate adversarial learning based data augmentation for disordered speech recognition tasks. 
In contrast, previous application of GANs were primarily studied in normal speech task domains including 
speech enhancement \cite{pascual2017segan, phan2020improving}, 
speaker verification \cite{DBLP:conf/interspeech/MichelsantiT17}, 
speech synthesis and voice conversion \cite{saito2017statistical, kameoka2018stargan}, 
code-switching sentence generation \cite{DBLP:conf/interspeech/ChangCL19}
and speech emotion recognition \cite{DBLP:conf/icassp/ChangS17}. 
Furthermore, previous researches on GANs for disordered speech processing were limited to impaired speech restoration \cite{chen2018generative} and classification \cite{jiao2018simulating}, but not for ASR system development as considered in this paper. 
Finally, the final system constructed using the best adversarial augmentation approach in this paper produced the lowest WER of 25.89\% on the UASpeech task published so far.

The rest of this paper is organized as follows. 
Conventional data augmentation techniques based on tempo, speed are presented in Section 2.
Section 3 proposes GAN based disordered speech data augmentation. 
Section 4 presents experiments and results on the UASpeech database. 
The last section concludes and discusses possible future works. 

\section{Disordered Speech Data Augmentation}

\begin{figure}
    \centering
  	\subfigure[CTRL]{
		\includegraphics[height=33pt]{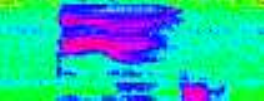}
	    \label{fig:subfig:ctrl}
	  }
	  \subfigure[DYS]{
		\includegraphics[height=33pt]{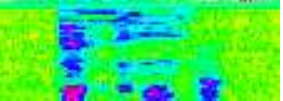}
	    \label{fig:subfig:dys}
	  }
	  
	  \subfigure[Tempo]{
    	\includegraphics[height=33pt]{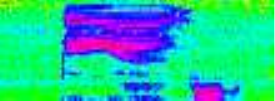}
	    \label{fig:subfig:tempo_ctrl}
	  }
	  \subfigure[Speed]{
    	\includegraphics[height=33pt]{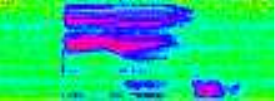}
	    \label{fig:subfig:speed_ctrl}
	  }
	  
	  \subfigure[tempo-GAN]{
    	\includegraphics[height=33pt]{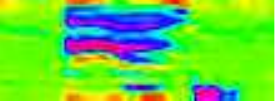}
	    \label{fig:subfig:tempo_gan}
	  }
	  \subfigure[speed-GAN]{
    	\includegraphics[height=33pt]{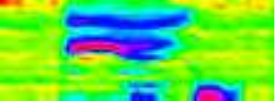}
	    \label{fig:subfig:speed_gan}
	  }
  \vspace{-4mm}
  \caption{Example spectrogram of (a) control, (b) dysarthric, (c) tempo or (d) speed perturbed control speech, (e) tempo-GAN or (f) speed-GAN generated ``dysarthric" speech.}
  \label{fig:ctrl_and_dys_fbank}
  \vspace{-6mm}
\end{figure}

In this section, we present two traditional data augmentation methods, tempo perturbation and speed perturbation for disordered speech recognition.
These serve as the baseline augmentation approaches modifying the overall speaking rate, volume and spectral shape, and the necessary time alignment between normal and dysarthric speech utterances to facilitate GAN based augmentation model training in Section 3. 

\noindent
\textbf{2.1. Tempo Perturbation:}
Tempo perturbation modifies the duration of the input time-domain signal $x(t)$, while keeping the overall shape of its spectrum untouched \cite{verhelst1993overlap, kanda2013elastic}.
This is first implemented by decomposing $x(t)$ into short analysis blocks.
These blocks are relocated along the time axis to construct the perturbed output $y(t)$.
This can be obtained using the waveform similarity overlap-add (WSOLA) algorithm \cite{verhelst1993overlap}, which maximizes the similarity between $x(t)$ and $y(t)$ by finding the optimal position of each analysis block iteratively.

Suppose the time-domain audio signal $x(t)$ is decomposed into short analysis blocks $\tilde{x}_m(r)$.
Given the analysis hopsize $H_\alpha$, these blocks are equally spaced along the time axis.
The synthesis blocks $\tilde{y}(r)$ are relocated along the time axis given the synthesis hopsize $H_s$ and perturbation factor $\alpha$ as:
\begin{equation}
	\setlength{\abovedisplayskip}{0.5pt}
	\setlength{\belowdisplayskip}{0.5pt}
	H_s = \alpha \times H_\alpha
\end{equation}
WSOLA iteratively updates the positions of $\tilde{x}_m(r)$.
For each $\tilde{x}_m(r)$, its center is shifted by $\Delta_m \in [-\Delta_{max}, \Delta_{max}]$ along the time axis.
By maximizing the cross-correlation between $\tilde{x}_m(r)$. and $\tilde{x}_{m-1}(r)$, WSOLA obtains the optimal value of $\Delta_m$.
This method guarantees that when both frames use the synthesis hopsize $H_s$, the optimal alignment can be achieved between the periodic structures of the perturbed analysis blocks and those of the previously copied synthesis blocks in the overlapping region.
The Hann window function $w(r)$ is then applied on the perturbed analysis block to compute the synthesis block $\tilde{y}_m(r)$ \cite{verhelst1993overlap}.
After finishing all iterations, the synthesis blocks are processed to reconstruct the actual time-domain perturbed output signal $y(t)$ using overlap and add (OLA) \cite{1162950}.
The perturbed signal $y(t)$ has a different duration, for example, representing a slower speaking rate of an impaired speaker, but keeps the same overall spectral shape as $x(t)$.

\begin{figure*}[!h]
  \centering
  \scalebox{0.9}{
	\subfigure[Training stage] {
  		\includegraphics[height=105pt]{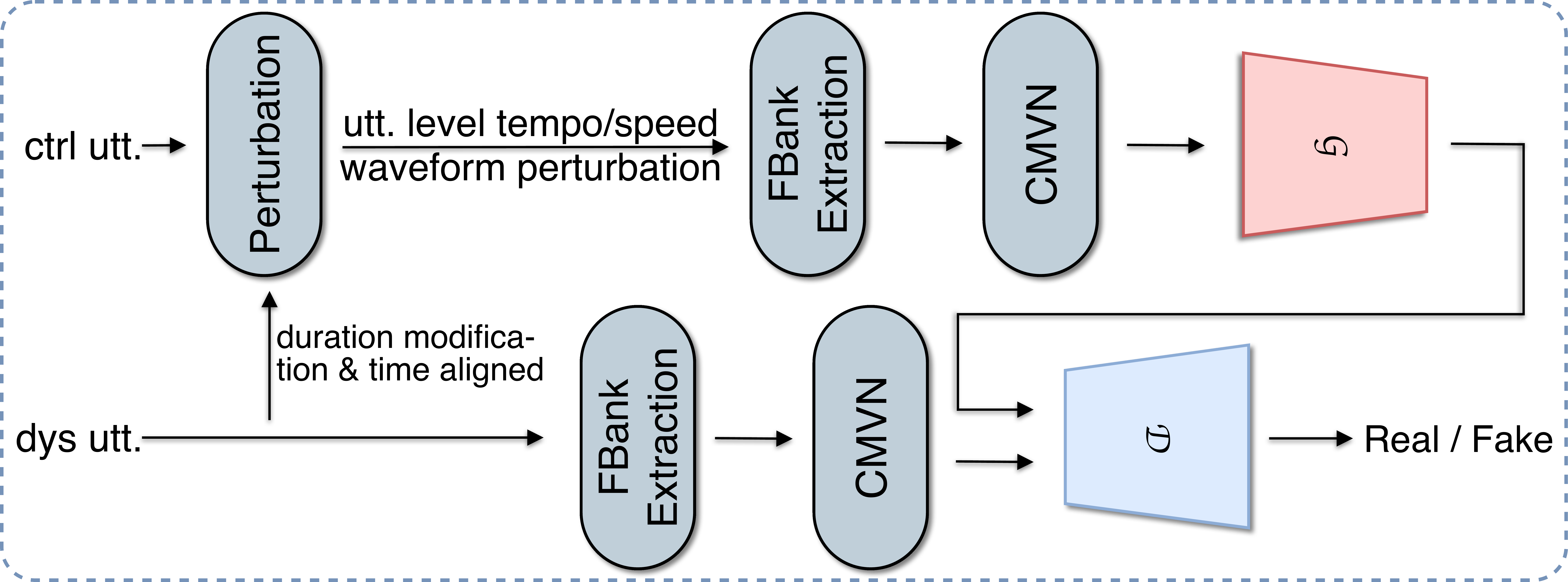}
  		\label{fig:subfig:training}
  	}
  	\subfigure[Generation stage] {
  		\includegraphics[height=105pt]{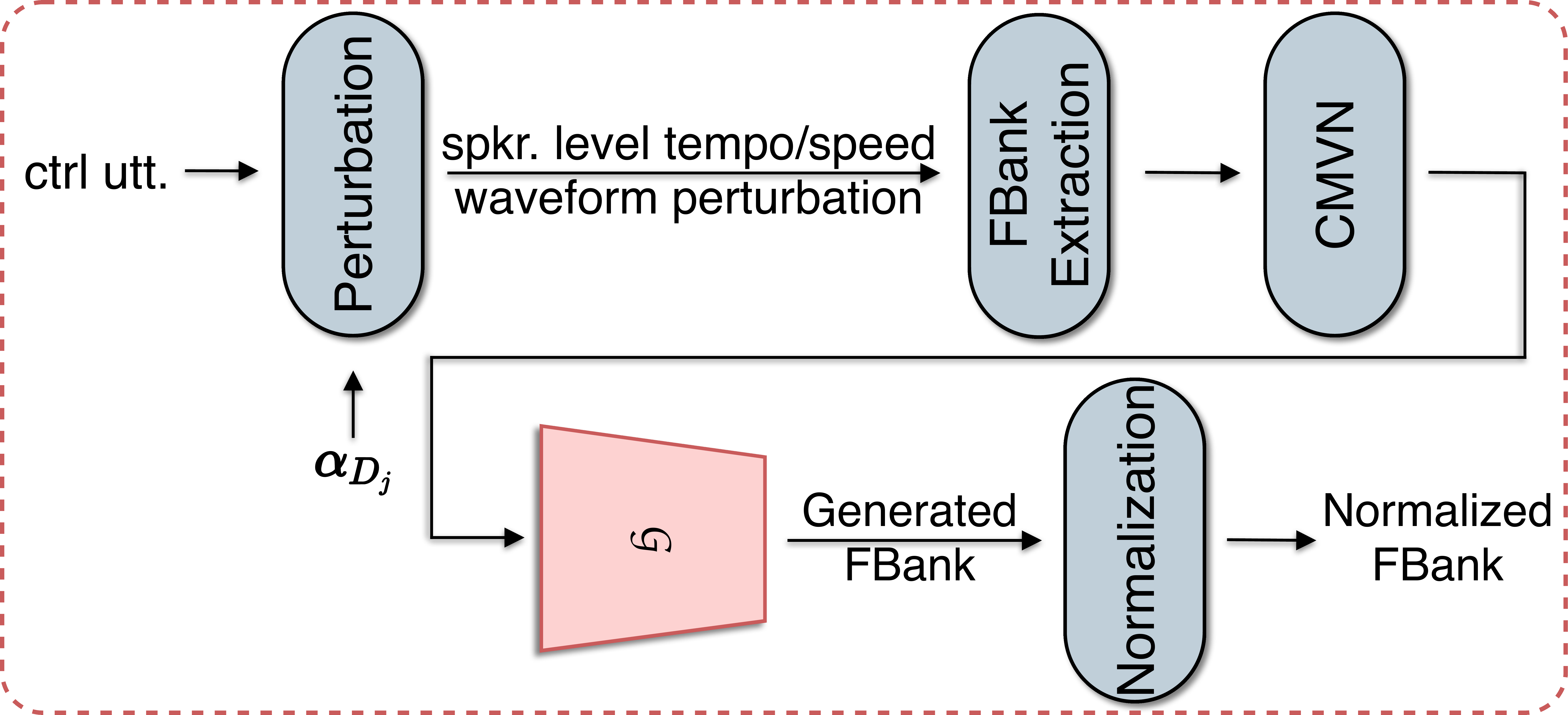}
  		\label{fig:subfig:conversion}
  	}
  }
  \vspace{-3.5mm}
  \caption{Illustration of (a) GAN model training on control and dysarthric utterances with modified duration and time alignment; and (b) GAN based dysarthric speech generation using disordered speaker level tempo/speed perturbed normal speech. Here $\alpha_{D_j}$ stands for the speaker level perturbation factor described in Section 2.3.}
  \vspace{-6.5mm}
  \label{fig:training}
\end{figure*}

\noindent
\textbf{2.2. Speed Perturbation:}
Speed perturbation \cite{ko2015audio} modifies the input time domain speech signal $x(t)$ by scaling the sampling resolution via a perturbation factor $\alpha$. 
The resulting speed perturbed signal output $y(t)$ is given as:
\begin{equation}
	\setlength{\abovedisplayskip}{0.5pt}
	\setlength{\belowdisplayskip}{0.5pt}
	y(t)=x(\alpha t)
\end{equation}
The above time-domain signal modification is equivalent to the following performing in the frequency domain:
\begin{equation}
	\setlength{\abovedisplayskip}{0.5pt}
	\setlength{\belowdisplayskip}{0.5pt}
	X(f) \rightarrow \frac{1}{\alpha}X(\frac{1}{\alpha}f)
\end{equation}
where $X(f)$ and $\frac{1}{\alpha}X(\frac{1}{\alpha}f)$ denote the Fourier transform of $x(t)$ and $y(t)$ respectively.
Speed perturbation changes the audio duration and overall spectral shape at the same time \cite{ko2015audio} to simulate, for example, a slower speaking rate, changes in speech volume and formant position of an impaired speaker.

\noindent
\textbf{2.3. Speaker Dependent Data Augmentation:}
During disordered speech augmentation, a combined use of speaker independent, global tempo or speed perturbation factors applied to disordered speech, for example, $\{0.9, 1.0, 1.1\}$, in common with the data perturbation methods widely used in normal ASR tasks \cite{panayotov2015librispeech}, and impaired speaker dependent (SD) perturbation of normal speech can be used to obtain the best performance \cite{geng2020investigation}. 
The disordered SD perturbation factors were obtained based on phonetic alignment and duration analysis described in \cite{xiong2019phonetic}.
We perform force alignment using a GMM-HMM system to get frame-level phoneme alignments, and then calculate the average phoneme duration.
Suppose the phoneme duration for control speaker $C_i$ and dysarthric speaker $D_j$ are denoted by $l_{C_i}$ and $l_{D_j}$.
For dysarthric speaker $D_j$, we take the average of $l_{C_i}$, denoted by $l_{\overline{C}}$, the SD perturbation factor $\alpha_{D_j}$ is given by:
\begin{equation}
	\setlength{\abovedisplayskip}{0.5pt}
	\setlength{\belowdisplayskip}{0.5pt}
	\alpha_{D_j}=\frac{l_{\overline{C}}}{l_{D_j}}
\end{equation}
$\alpha_{D_j}$ is then used as the perturbation factor when performing data augmentation using normal speech for speaker $D_j$.

\section{GAN Based Data Augmentation}

\begin{table*}[!ht]
  \caption{Performance on the 16 UASpeech dysarthric speakers of different data augmentation methods. ``CTRL" and ``DYS" refer to control and disordered speech. ``Tmp." and ``Spd." stand for tempo and speed perturbation, subscript ``$_G$" represents GAN based augmentation. ``1x" and ``2x" stand for the amount of augmented data. ``2x" for disordered speech represents speed perturbation using global perturbation factors 0.9 and 1.1. ``VL", ``L", ``M" and ``H" denote intelligibility levels (Very Low / Low / Medium / High). ``$\dag$" denotes a statis. sig. improvement$^2$ is obtained from Sys. 3-5, Sys. 7-9, Sys. 11-19 against Sys. 2, 6, 10 respectively.}
  \vspace{-3mm}
  \label{tab:experiments_of_sys}
  \centering
  \renewcommand\tabcolsep{6.5pt}
  \scalebox{0.788}{
	\begin{tabular}{c|c|c|c|c|c|c|c|c|c|c|c||c|c|c|c|c}
	    \hline
	    \hline
		\multirow{3}{*}{Sys.} & \multicolumn{5}{c|}{Data Augmentation} & \multirow{3}{*}{Hrs.} & \multicolumn{5}{c||}{WER \% (Unadapted)} & \multicolumn{5}{c}{WER \% (LHUC-SAT Adapted)} \\
		\cline{2-6}
		\cline{8-17}
		& \multicolumn{4}{c|}{CTRL} & DYS &  & \multirow{2}{*}{VL} & \multirow{2}{*}{L} & \multirow{2}{*}{M} & \multirow{2}{*}{H} & \multirow{2}{*}{O.A.} & \multirow{2}{*}{VL} & \multirow{2}{*}{L} & \multirow{2}{*}{M} & \multirow{2}{*}{H} & \multirow{2}{*}{O.A.} \\
		\cline{2-6}
		& Tmp. & Tmp.$_G$ & Spd. & Spd.$_G$ & Spd. &  &  &  &  &  &  &  & &  &  &  \\
		\hline
		\hline
		1  & \multicolumn{5}{c|}{NA} & 30.6 & 69.82 & 32.61 & 24.53 & 10.40 & 31.45 & 64.39 & 29.88 & 20.27 & 8.95 & 28.29 \\
		\hline
		\hline
		2  & 1x & -  & -  & -  & \multirow{4}{*}{-} & \multirow{4}{*}{50.2} & 69.76 & 33.20 & 25.17 & 10.59 & 31.79 & 68.32 & 29.65 & 21.09 & 8.98 & 29.24			\\
		3  & -  & 1x & -  & -  &   &   & 69.35 & 31.89 & 22.31 & 9.87 & 30.57$^\dag$ & 66.02 & 28.20 & 19.41 & 8.11 & 27.76$^\dag$ \\
		4  & -  & -  & 1x & -  &   &  & 67.52 & 31.55 & 21.96 & 9.57 & \textbf{29.92}$^\dag$ & 65.27 & 28.74 & 19.78 & 8.26 & 27.86$^\dag$ \\
		5  & -  & -  & -  & 1x &   &   & 69.78 & 32.98 & 23.88 & 10.36 & 31.41 & \textbf{64.54} & 28.27 & 18.82 & 8.41 & \textbf{27.45}$^\dag$ \\
		\hline
		\hline
		6 & 1x & -  & -  & -  & \multirow{4}{*}{2x} & \multirow{4}{*}{87.5} & 66.18 & 31.48 & 22.29 & 9.82 & 29.77 & 62.31 & 27.24 & 18.43 & 8.50 & 26.66 \\
		7 & -  & 1x & -  & -  &   &   & 66.95 & 29.20 & 20.13 & 10.00 & 28.99$^\dag$ & 61.14 & 26.95 & 18.01 & 8.30 & \textbf{26.19}$^\dag$ \\
		8 & -  & -  & 1x & -  &   &   & 66.48 & 29.80 & 21.88 & 9.55 & 29.23$^\dag$ & 61.10 & 27.88 & 18.37 & 7.97 & 26.38 \\
		9 & -  & -  & -  & 1x &   &   & 64.45 & 29.19 & 21.47 & 9.75 & \textbf{28.63}$^\dag$  & \textbf{60.83} & 27.34 & 18.35 & 8.11 & 26.23 \\
		\hline
		\hline
		10 & 2x & -  & -  & -  & \multirow{4}{*}{2x} & \multirow{4}{*}{130.1} & 68.05 & 32.60 & 23.19 & 9.81 & 30.63 & 61.94 & 28.27 & 19.54 & 8.12 & 26.94 \\
		11 & -  & 2x & -  & -  &   &   & 65.73 & 29.96 & 20.62 & 9.53 & \textbf{28.86}$^\dag$ & 62.51 & 28.19 & 17.37 & 8.10 & 26.61	 \\
		12 & -  & -  & 2x & -  &   &   & 66.45 & 28.95 & 20.37 & 9.62 & \textbf{28.73}$^\dag$ & 62.50 & 27.26 & 18.41 & 8.04 & 26.55 \\
		13 & -  & -  & -  & 2x &   &   & 66.14 & 29.65 & 20.62 & 9.94 & 29.01$^\dag$  & \textbf{61.42} & 27.37 & 16.50 & 7.75 & \textbf{25.89}$^\dag$ \\
		\hline
		\hline
		14 & - & 1x & 1x & -  & \multirow{2}{*}{2x} & \multirow{2}{*}{109.1} & 65.34 & 29.15 & 20.92 & 9.77 & 28.70$^\dag$ & \textbf{61.10} & 26.78 & 17.60 & 7.80 & \textbf{25.89}$^\dag$ \\
		15 & -  & - & 1x  & 1x  &   &   & 65.22 & 29.22 & 19.82 & 9.52 & \textbf{28.40}$^\dag$ & 61.49 & 27.93 & 17.60 & 8.02 & 26.35 \\
		\hline
		\hline
		16 & - & 2x & 2x & -  & \multirow{2}{*}{2x} & \multirow{2}{*}{194.3} & 65.56 & 30.07 & 20.64 & 9.45 & 28.83$^\dag$ & \textbf{60.99} & 27.97 & 17.54 & 8.03 & 26.25$^\dag$ \\
		17 & -  & - & 2x  & 2x  &   &   & 65.81 & 29.94 & 20.17 & 9.49 & \textbf{28.77}$^\dag$ & 61.37 & 27.66 & 17.43 & 7.73 & \textbf{26.12}$^\dag$ \\
		\hline
		\hline
		18 & - & 1x & - & 1x  & \multirow{2}{*}{2x} & 109.1 & 65.50 & 29.80 & 20.86 & 9.49 & \textbf{28.80}$^\dag$ & \textbf{60.69} & 27.33 & 17.98 & 8.02 & \textbf{26.09}$^\dag$	 \\
		19 & -  & 2x & -  & 2x  &   &  194.3 & 67.93 & 30.60 & 21.11 & 9.53 & 29.59$^\dag$ & 61.78 & 27.52 & 17.13 & 7.83 & 26.15$^\dag$ \\
		\hline
		\hline
  	\end{tabular}
  }
  \vspace{-6.5mm}
\end{table*}

As discussed in Section 1, conventional tempo or speed perturbation of Section 2 only characterize an overall decrease in speaking rate, speech volume and changes in spectral envelope found in disordered speech. 
In this section, DCGANs are proposed to further inject other prominent features associated with disordered speech such as articulatory imprecision, breathy and hoarse voice, into tempo or speed perturbed normal speech Filter Bank (FBank). 
For instance, in contrast to the example healthy speech segment’s FBank of word ``burrows" in Fig.\ref{fig:subfig:ctrl}, the comparable dysarthric speech spectrogram in Fig.\ref{fig:subfig:dys} contains not only incorrectly weakened formants indicating articulatory imprecision, but also some energy distributed across higher frequencies at both the start and end of the utterance due to possible difficulty in breath control when speaking.


\begin{figure}
	\centering
	\includegraphics[width=0.75\linewidth]{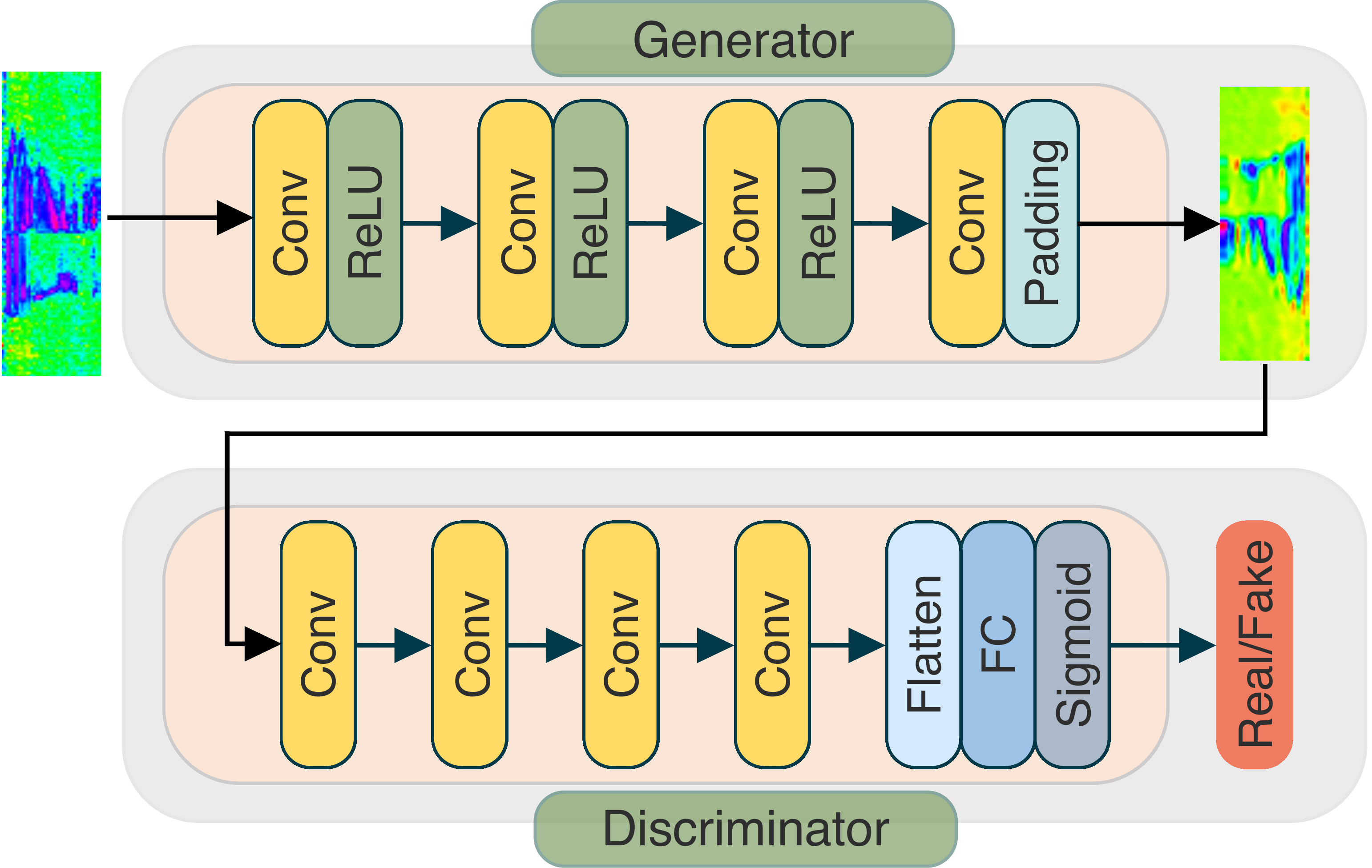}
	\vspace{-3mm}
	\caption{The architecture of the proposed GAN.}
	\label{fig:gan_architecture}
	\vspace{-7mm}
\end{figure}

\noindent
\textbf{3.1. GAN Architecture:}
The overall architecture configurations of the proposed DCGAN model follow those used in \cite{jiao2018simulating}, also again shown in Fig.\ref{fig:gan_architecture}.
The Generator component contains 4 convolutional layers, the first three of which have 8 kernels while the last one has 1 kernel only. 
All of the kernels in the Generator has a kernel size of $3 \times 3$ and stride of $1 \times 1$. 
Each of the first three convolutional layers is also immediately connected to ReLU activations. 
We use Replicate Padding to replicate the edges of the feature map to ensure that the output and input dimensions are the same.
The Discriminator component contains 4 convolutional layers of 8, 16, 32 and 64 kernels respectively, all of which use a kernel size of $2 \times 2$ and stride of $2 \times 2$. 
A flattening operation is applied to concatenate the outputs of convolutional layers, resulting in a 3000-dimensional vector.
A fully connected layer with Sigmoid activation is used for binary classification in the Discriminator.

\noindent
\textbf{3.2. GAN Training:}
Prior to GAN model training, pairs of normal and disordered speech utterances of identical word contents but often different lengths need to be formed. 
In order to facilitate a frame-by-frame comparison between the GAN perturbed normal speech spectrogram against that of the target disordered speech, each normal speech segment is either tempo or speed perturbed to produce a modified duration that matches against that of a target disordered speech utterance, as shown in Fig.\ref{fig:subfig:training}. 
This requires a scaling factor $\alpha$ to be estimated for each such normal and disordered speech segment pair using phone alignment analysis similar to the procedure described in Section 2.3 for speaker level perturbation. 
The resulting pairs of normal and disordered speech utterances that now have the same durations are further zero mean and unit variance normalized at speaker level, before being used in GAN model training.  

The GAN training objective function that both maximizes the binary classification accuracy on target disordered speech and minimizes that obtained on the GAN perturbed normal speech, in the hope that upon convergence the latter is modified to be sufficiently close to the target impaired speech, is given by
\begin{equation}
	\setlength{\abovedisplayskip}{0.5pt}
	\setlength{\belowdisplayskip}{0.5pt}
	\begin{aligned}
		\mathop{min}\limits_{G_j}^{}\mathop{max}\limits_{D_j}^{} &\ V(D_j, G_j) \\
		&\ = \mathbb{E}_{f_{D}\sim p_{D_j}(f)}[\log{(D_j(f_{D_j}))}] \\
		&\ + \mathbb{E}_{f_{C} \sim p_{C}(f)}[\log{(1-D_j(G_j(f_{C})))}]
	\end{aligned}
\end{equation}
where $j$ represents the index for target dysarthric speaker, $G_j$ and $D_j$ are Generator and Discriminator associate with dysarthric speaker $j$, $f_C$ and $f_{D_j}$ stand for the FBank features of paired control and dysarthric utterances.
During GAN model training for each target impaired speaker, the learning rate for both the Generator and Discriminator components is halved every 2500 iterations until convergence.

\noindent
\textbf{3.3. Disordered Speech Spectrum Generation:}
During GAN based data augmentation, as shown in Fig.\ref{fig:subfig:conversion}, the target disordered speaker level tempo or speed perturbed 40-dimensional FBank features obtained from normal speech using the two baseline approaches described in Section 2.3 are fed into the corresponding Generator components to produce the comparable speaker level ``tempo-GAN" or ``speed-GAN" (``Tmp.$_G$" or ``Spd.$_G$" in short in Table \ref{tab:experiments_of_sys}) augmented disordered data, before being further zero mean and unit variance normalized for speech recognition system development.

\section{Experiments and Results}


All experiments are conducted on UASpeech database, which is a dataset for the single word recognition task.
It contains 102.7 hours of speech from 16 dysarthric and 13 control speakers with 155 common words and 300 uncommon words.
Utterances are divided into 3 blocks, each containing all common words and one-third of the uncommon words.
Block 1 and 3 are treated as training set while block 2 of the 16 dysarthric speakers is treated as test set.
In the preprocessing stage, we use a GMM-HMM system to implement silence stripping as described in our previous work \cite{liu2020exploiting}.
Without data augmentation, the final training set contains 99195 utterances, around 30.6 hours.
The test set contains 26520 utterances, around 9 hours.

\noindent
\textbf{4.1. Experiment Setup:}
The proposed GAN model is implemented with PyTorch \cite{NEURIPS2019_9015}.
The ASR System is implemented using an extended Kaldi toolkit \cite{povey2011kaldi}, featuring the same architecture in our previous work \cite{geng2020investigation}.
We use HTK toolkit \cite{young2002htk} for phonetic analysis, silence stripping and feature extraction.
Both tempo and speed perturbation are implemented using SoX$^1$ \footnote{$^1$ Sox, audio manipulation tool. Available: http://sox.sourceforge.net}.

\noindent
\textbf{4.2. Performance of Data Augmentation:}
Table \ref{tab:experiments_of_sys} (col.8-12) presents the performance of unadapted systems with different data augmentation methods prior to applying LHUC-SAT (last five columns). 
We can observe several trends from Table \ref{tab:experiments_of_sys}: 
1) Our proposed tempo-GAN approach consistently outperforms the ones with tempo perturbation only (Sys. 3 \textit{vs.} 2, Sys. 7 \textit{vs.} 6 and Sys. 11 \textit{vs.} 10, col. 12) by up to 1.77\% absolute (5.77\% relative) overall WER reduction (Sys. 11 \textit{vs.} 10, col. 12);
2) Our proposed speed-GAN approach outperform the ones with speed perturbation only (Sys. 9 \textit{vs.} 8, col. 12) by 0.6\% absolute (2.05\% relative) overall WER reduction;
3) Col. 12 of Sys. 14-19 explore different combinations of GAN generated data and tempo/speed perturbed speech. Sys. 15 gives the largest WER reduction than the baseline system.

\footnote{$^2$ A matched pairs sentence-segment word error based statistical significance test was performed at a significance level $\alpha=0.05$.}

\begin{table}[h]
	\centering
	\caption{A comparison between published systems on UASpeech and our system. ``DA" stands for data augmentation.}
	\vspace{-3mm}
    \renewcommand\tabcolsep{0pt}
    \label{tab:comparisons}
    \scalebox{0.75}{
	\begin{tabular}{cc}
		\toprule
		Systems & WER \% \\
		\hline
		Sheffield-2013 Cross domain augmentation \cite{christensen2013combining} & 37.50 \\
		Sheffield-2015 Speaker adaptive training \cite{sehgal2015model} & 34.80 \\
		CUHK-2018 DNN System Combination \cite{yu2018development} & 30.60 \\
		Sheffield-2019 Kaldi TDNN + DA \cite{xiong2019phonetic} & 27.88 \\
		Sheffield-2020 Fine-tuning CNN-TDNN speaker adaptation \cite{xiong2020source} & 30.76 \\
		CUHK-2020 DNN + DA + LHUC SAT \cite{geng2020investigation} & 26.37 \\
		CUHK-2021 LAS + CTC + Meta Learning + SAT \cite{wang2021improved}
                        & 35.00 \\
		\textbf{DNN + GAN based DA + LHUC SAT (Table \ref{tab:experiments_of_sys}, Sys. 13 \& 14)} & \textbf{25.89} \\
		\toprule
	\end{tabular}
	}
	\vspace{-7.9mm}
\end{table}

\noindent
\textbf{4.3. Performance after LHUC-SAT:}
To model the variability in both original and augmented data, we perform LHUC based speaker adaptive training on the systems.
With LHUC-SAT, trends can be observed:
1) LHUC-SAT can bring up to 3.96\% (Sys. 5, last col.) absolute (12.61\% relative) WER reduction compared with those without adaptation (Sys. 5, col. 12);
2) When further combined with LHUC-SAT, the proposed GAN based approaches consistently outperform the ones with perturbation only (Sys. 2-13, last col.) by up to 1.48\% absolute (Sys. 3 \textit{vs.} 2, last col.) (5.06\% relative) WER reduction;
3) With LHUC-SAT, the tempo and speed GAN augmented systems 13 \& 14 give the lowest WER of 25.89\% among all systems in Table \ref{tab:experiments_of_sys}. 
This is also the best WER performance reported on UASpeech as summarised in Table \ref{tab:comparisons}.

\section{Conclusions}
This paper presents GAN based data augmentation approaches for disordered speech recognition. 
Experimental results suggest improved coverage in augmented data and model generalisation were obtained over baseline systems using no augmentation and conventional methods based on tempo or speed perturbation. 
Future research will focus on improve GAN modelling on richer spectral and temporal characteristics of disordered speech.

\section{Acknowledgements}
This research is supported by Hong Kong RGC GRF grant No. 14200218, 14200220, TRS T45-407/19N, ITF grant No. ITS/254/19, and SHIAE grant No. MMT-p1-19.

\bibliographystyle{IEEEtran}

\bibliography{mybib}

\end{document}